# Terahertz All-Optical Modulation in a Silicon-Polymer Hybrid System

Michael Hochberg,[1*] Tom Baehr-Jones,[1*] Guangxi Wang,[1] Michael Shearn,[1] Katherine Harvard,[1] Jingdong Liu,[2] Baoquan Chen,[2] Zhengwei Shi,[2] Rhys Lawson,[3] Phil Sullivan,[3] Alex K. Y. Jen,[2] Larry Dalton,[3] Axel Scherer[1]

[1]*Department of Applied Physics, 1200 E California Blvd., California Institute of Technology, Pasadena California 91125*

[2]*Department of Materials Science and Engineering, University of Washington, Seattle, Washington, 98195*

[3]*Department of Chemistry, University of Washington, Seattle, Washington, 98195*

\* These authors each contributed equally to this work. Address correspondence to Hochberg@caltech.edu (M. H.), or to thorolf@caltech.edu (T. B.-J.)

**Although Gigahertz-scale free-carrier modulators have been previously demonstrated in silicon, intensity modulators operating at Terahertz speeds have not been reported because of silicon's weak ultrafast optical nonlinearity. We have demonstrated intensity modulation of light with light in a silicon-polymer integrated waveguide device, based on the all-optical Kerr effect — the same ultrafast effect used in four-wave mixing. Direct measurements of time-domain intensity modulation are made at speeds of 10 GHz. We showed experimentally that the ultrafast mechanism of this modulation functions at the optical frequency through spectral measurements, and that intensity modulation at frequencies in excess of 1 THz can be obtained in this device. By integrating optical polymers through evanescent coupling to high-mode-confinement silicon waveguides, we greatly increase the effective nonlinearity of the waveguide for cross-phase modulation. The combination of high mode confinement, multiple integrated**

**optical components, and high nonlinearities produces all-optical ultrafast devices operating at continuous-wave power levels compatible with telecommunication systems. Although far from commercial radio frequency optical modulator standards in terms of extinction, these devices are a first step in development of large-scale integrated ultrafast optical logic in silicon, and are two orders of magnitude faster than previously reported silicon devices.**

If we are to build all-optical computers, as well as practical high-bandwidth all-optical wavelength converters and modulators, we must have the ability to modulate light with light in a chip-scale, highly integrated platform[1,2,3,4]. The ability to pattern nanoscale structures at wafer scales, the existence of commercial electronic chip fabrication processes, and the low optical loss and high mode confinement available in silicon waveguides all combine to make silicon-on-insulator an attractive material system for very large-scale integrated photonics[5,6,7]. Recently, compact electrooptic modulators based on free-carrier effects operating at speeds up to 10 GHz,[8,9,10,11,12] electrooptic polymer-based switches and optical rectification-based detectors,[13], lasers[14,15,16], and stimulated-Raman-scattering–based modulators[17] have all been demonstrated in silicon. However, integrated, low-power, ultrafast devices operating with speeds in the Terahertz have not previously been realized.

Here we report experimental demonstration of intensity modulation of light with light in a planar, hybrid silicon-on-insulator–polymer, all-optical modulator. This device is based on the optical Kerr effect, which underlies effects such as ultrafast four-wave mixing. We show that similar devices could operate at speeds up to 1 Terahertz. The optical nonlinearity is substantially enhanced by a cladding of engineered $\chi^3$ nonlinear optical polymer.[18,19] The modulator is operated by off-the-shelf continuous-wave lasers and erbium-doped fiber amplifiers, with intensity levels of 100 mW per laser at the entrance to the gate (but after the fiber-to-chip coupler) and of around 30 mW at the source input port.  Such power levels are common in today's optical telecommunications systems.[20] Our device is based on a Mach–Zehnder geometry, as is shown in figure 1. The source waveguide is split into two arms; in one of the arms, a gate signal is introduced via a 3dB coupler. The nonlinear Kerr effect enables the gate signal to induce a phase shift in the source signal, which in turn allows the shifted source signal to interfere with the optical signal that has travelled down a reference arm at the end of the Mach–Zehnder interferometer, causing an intensity modulation of the source signal. The Mach–Zehnder interferometer is unbalanced, which allows us to control the intrinsic phase shift between the arms by tuning the source wavelength.

The precise behaviour of the entire nonlinear system is complex: Not only is there a phase shift in the source signal, but also the four-wave mixing process produces sidebands from both the source and the gate laser, which must be considered in analyzing the behaviour of the device. In the low-conversion regime, only a small fraction of the source wavelength is converted into other wavelengths via four-wave mixing, and the primary effect of the four-wave mixing process is to phase shift the signal wave by a small amount. It is straightforward to show the effect of the gate signal on the source signal in this regime. Let $w_s$, $w_g$ and $a_s$, $a_g$ be the frequencies and amplitudes of the source and gate lasers, respectively, and the sideband $w_a = w_s + (w_s - w_g)$. Then,

$$a_s(L) = a_s(0)\exp(\frac{i6w_s\chi^3}{2cn}|a_g(0)|^2 L) \tag{1}$$

$$a_a(L) = \frac{i3\omega_a \chi^3}{2cn} a_s^2(0) a^*_g(0) \int_0^L \exp(-\frac{2}{c}\frac{\partial n}{\partial w}(w_s - w_g)^2 zi)dz \qquad (2)$$

Here, we have assumed that we can neglect second- and higher-order derivatives of the waveguide effective index, which is reasonable for our waveguide geometry. Similar expressions exist for the other sideband at $w_g+(w_g-w_s)$, and for the effect on $a_g$. The crucial point is that, in the low-conversion regime, the effect of the gate signal is to phase shift the source signal. The power level of the source signal is unchanged. Note also that the dispersion of the waveguide does not affect this phase shift. Thus, the bandwidth of the device is limited solely by the effect of dispersion on the intensity-modulation envelope of the gate signal as the latter travels through the waveguide. The dispersive properties of the waveguide used for this device are well known from both theory and experiment, and it is straightforward to show that, for a device of 1-cm length, minimal pulse distortion in the gate will occur up to intensity modulations of 1 THz (see *Supporting Online Materials,*).

The most straightforward method of observing the intensity modulation is simply to send in a modulated gate laser and an unmodulated source laser, and to observe the intensity modulation on the source wavelength. We initially exercised each device with a swept-wavelength laser and diode detector, to determine the optical loss and wavelength response between the gate and drop port, and between the source and drop ports. The devices were then connected to an Agilent 8703B, which is a radio-frequency vector-network analyzer operating both in the electrical and in the optical domain. A single laser, modulated with a sine wave at frequencies between 50 MHz and 10 GHz, was run through an EDFA and a polarization controller, and used as the gate. The source was provided by a diode laser, which was amplified by a separate polarization-maintaining EDFA. At the input to the phase modulation arm, the power in the source frequency was 10 dBm; that in the gate was 17 dBm. The output of the modulator was either coupled into an optical spectrum analyzer (Figs. 2C and 2D) or was filtered to suppress the gate wavelength, and then fed into the input port of the vector network analyzer (Fig. 2B). The optical input port of the vector-network analyzer is a 20 GHz bandwidth photodiode. The Agilent 8703B is designed to measure the transfer of an intensity modulation from one electrical- or laser-signal port to another such port. Such a system is typically used to measure the performance of modulators and detectors as a function of frequency. It reports the results of the measurement as an optical S parameter, which is defined as follows. Let I(t) be time domain intensity measured by the photodiode. Then, the optical S parameter $S$ is defined in dBo as

$$S(f) = 10\log 10(|\int_0^T \exp(i2\pi\, ft)I(t)dt\,|) + N \qquad (3)$$

Here, $T$ is an integration period for the measurement, and $f$ is the frequency at which the optical $S$-parameter is being measured. $N$ is a normalization offset selected such that, if the gate laser was simply directed into the photodetector directly (a "through" measurement), the optical S parameter would be 0 dB. As shown in Figure 2, the optical S parameter for our nonlinear device was about 5 dB above the noise floor, asymptotically approaching −37 dB. This non-zero S-parameter indicates that there is a significant amount of intensity transfer from the gate beam to the signal beam; as we can see from Equation 3, a mere phase modulation of the signal would not be detected. In taking these measurements, we noted that the strength of the optical S-parameter depended strongly on the precise wavelength of the signal. This relation provides an indication that the intensity transfer is caused by the Mach–Zehnder geometry, as the sensitivity of an output of a Mach–Zehnder modulator to a phase shift in one arm is highly dependant on the relative phase shift between the two arms.

Three control experiments helped us to establish that these measurements reflect the phenomenon described. First, measurements were performed with the gate signal and pump signal turned off individually, and then with both turned off concurrently; in all three cases, no intensity coupling was observed, as detailed in Figure 2. Second, we ran experiments on devices that were clad with optically linear polymers, and on unclad devices, and observed no intensity modulation transfer in either case. Third, we built a device with just a single phase-modulation arm, and no Mach–Zehnder geometry. We saw no intensity modulation in this device either, although four-wave mixing peaks were observed, indicating phase modulation was occurring and that the third-order nonlinearity in these waveguides was comparable with that observed in the measured Mach–Zehnder devices. This experiment performed with the linear optical polymer shows that thermal and free-carrier mechanisms cannot be the source of the intensity modulation transfer. That the optical S parameter is stronger at lower frequencies is probably due to the nonlinear polymer having a higher $\chi^3$ moment for lower frequencies.

The most direct method to demonstrate that the intensity modulator shown here could in fact function in the terahertz regime would be, of course, simply to increase the rate of the intensity modulation and the photodetector bandwidth, in which case, we believe, the optical S parameter would continue roughly at the asymptote that it appears to approach in figure 2. Equipment to generate intensity modulations at this frequency is not readily available, however, and there is no simple photodetector that is capable of performing with the fantastically short integration times needed to observe such a result directly.

Accordingly, we measured the terahertz performance of the device indirectly, by means of spectral data. To show that the modulators' intensity-modulation mechanism functions in the terahertz regime, we need only to show that the four-wave mixing process occurs with the expected amplitudes in our device. Then, Equation 1 gives the exact amount of phase shift that occurs. Figure 3 shows a block diagram of the experiment, as well as several spectral measurements. As might be expected from an ultrafast nonlinearity, the usual set of four-wave mixing peaks are observed. The magnitude of the sidebands shows that the nonlinear effect is not attenuated, even at these extremely high frequencies; other modulation mechanisms that are limited to the 10's of gigahertz, such as free-carrier modulation, would not show such spectra.

We used three, rather than two, lasers in this experiment, to provide a comprehensive demonstration of the operation of the device. In addition to demonstrating the ultrafast nonlinear optical effect, this experiment helped us to show a crucial point, which was the privileged nature of the signal wavelength: Figure 3 shows that, when the signal wavelength is tuned, the phase modulation from the unbalanced Mach–Zehnder modulator can be seen to produce an intensity modulation on the source wavelength, while the relevant sidebands are not altered in height. The results in Figure 3 show that we are observing terahertz-scale nonlinear effects, as well as Mach–Zehnder modulator behaviour, in the same measurement.

By far the most important datum that can be extracted from the spectral measurements is the strength of the $\chi^3$. Given this value and Equation 1, we can deduce the phase shift on the source signal based on a given intensity of the gate signal, from which we can derive the optical S parameter. As shown in the supporting materials, the most straightforward method of deriving this S parameter is to relate the phase shift in radians to the relative extinction of the four-wave mixing peaks:

$$\Delta\phi = 4\sqrt{f} \qquad (4)$$

For the modulator shown in Figure 3, typical values of f were –40 dB, indicating an approximate phase shift of 0.04 radians with 14 dBm of gate power, implying a relatively small extinction of around 0.3 dB for this modulator device in this circumstance. By trading increased insertion loss for increased modulation ratio (by changing the point of operation of the unbalanced Mach–Zehnder interferometer), we could achieve an extinction of 3 dB with –24 dB of intrinsic insertion loss in our device. However, because of the signal-to-noise limitations of our test setup, we were able to operate in only the low-loss/low-extinction regime. An effective nonlinear $\chi^3$ coefficient can be estimated as $1\times10^{-20}$ $(m/V)^2$, and the nonlinear coefficient of the polymer is approximately three times that value. Typical results for $\Delta\varphi$ ranged from 0.02 to 0.04 radians during testing for this gate power level. As can be seen in Figure 2, the implied optical S parameter from these observations is in close agreement with the directly measured S-parameter value at 10 GHz.

Since the spectral measurements predict the observed modulation due to a $\chi3$ moment, and the various control experiments have eliminated mechanisms such as free-carrier nonlinearities, we believe that we have demonstrated that the modulation mechanism is an ultrafast optical nonlinearity. From equation 1, it is clear that the relative phases between the source and gate beam do not matter; it is only the overlap of the intensity of the gate beam that alters the phase of the source beam. Therefore, the bandwidth limit of our device will depend on the dispersion of the Silicon waveguides. Through the use of an eigenvalue solver, and confirmed via FDTD and experiment, we predict our waveguides to have a group velocity dispersion of approximately $-2000$ ps nm$^{-1}$ km$^{-1}$. This value is typical for comparable waveguides[21]. It is a large value when compared to the group velocity dispersion of large fiber-optical modes; however, its extremely small size makes our device less sensitive to the effects of dispersion than are fibers, which are often kilometres in length. In the supporting online materials, we show simulations in the time domain indicating that such dispersion is unlikely to be a limiting factor for devices such as the ones shown, up to speeds of at least 1 THz.

Our thoughts about the practical application of the results are based on several considerations. The observed phase shift is fairly low, whereas it is desirable in most Mach–Zehnder modulators to have at least $\pi$ radians of phase shift. However, polymer strengths have been consistently improving over the past several years, and we can reasonably expect this trend to continue. Also, it is very likely that we will be able to decrease the waveguide loss from the approximately 7 dB/cm of loss achieved in these devices to less than 1 dB/cm of loss[22]. In that case, we can increase the length of the device dramatically, although that change will in turn increase the limitations on bandwidth due to waveguide dispersion. If the nonlinear polymer were to increase in strength by a factor of 10, and the device were made 10 times longer, the phase shift achieved with the modest powers we have used would be about $\pi$. We believe it probable that such improvements can be made, rendering this device practical as a high-extinction and high-bandwidth all-optical modulator. Furthermore, resonant effects can be used to enhance the nonlinearity of these devices.

One practical device that could be built based on this mechanism is a high-bandwidth wavelength converter. That is, we could design a device that would take an intensity modulation at one wavelength, and transfer that modulation to another carrier wavelength. We can imagine a fiber-optical communications system in which a high-speed intensity modulation at a wavelength such as 1480 nm must be transferred to another wavelength at 1550 nm. A converter based on our nonlinear modulator would have several advantages. First, it would not require any high-speed electronics, as the conversion mechanism occurs entirely in the optical domain. Such a device would have an exceptionally high bandwidth. Of course, if the optical nonlinearity were strong enough to enable $\pi$ of phase shift, we could conceivably perform this conversion via four-wave mixing in a single straight waveguide. In that case, however, we would have to solve the problem of nearly exactly matching the effective index at two different wavelengths — a task that is extremely challenging in waveguides such as this. Even for pulse streams at 10 GHz, this phase matching condition would be just as serious a problem as at the much higher speeds. By contrast, our device, based on a Mach–Zehnder interferometer, does not have this limitation, and begins to suffer ill effects on performance from the dispersion of the waveguide only as switching speeds of 1 THz are approached. Such a device could be particularly desirable for low-latency applications, where the delay associated with an O-E and E-O conversion is unacceptable. For a 1-cm path-length device, we would expect delays on the scale of hundreds of picoseconds, whereas electronic E-O and O-E converters will typically have latency numbers of nanoseconds or more[23].

In conclusion, we have demonstrated ultrafast intensity modulation in a silicon-based device that takes advantage of the high modefield confinement in silicon waveguides, as well as the high nonlinearity of optical-polymer materials. As performances are improved, this type of device may find wide application in the data communications market, particularly in very high bandwidth/low-latency applications. Moreover, we believe that all-optical modulation of this nature is an important first step to all-optical logic and computation. We are able to measure the intensity modulation directly up to only the 10-GHz regime, due to equipment limitations; however, based on spectral measurements, we are confident that we have shown that an ultrafast nonlinearity is the mechanism of intensity-modulation transfer, and that the device functions as an intensity modulator into the terahertz.

**Acknowledgments**

We gratefully acknowledge research support from the National Science Foundation Center on Materials and Devices for Information Technology Research (CMDITR), through grant DMR-0120967. M.S. thanks the Department of Homeland Security for support of his graduate fellowship.  M. H. thanks the National Science Foundation for support of his graduate research fellowship.

**Figure Captions**

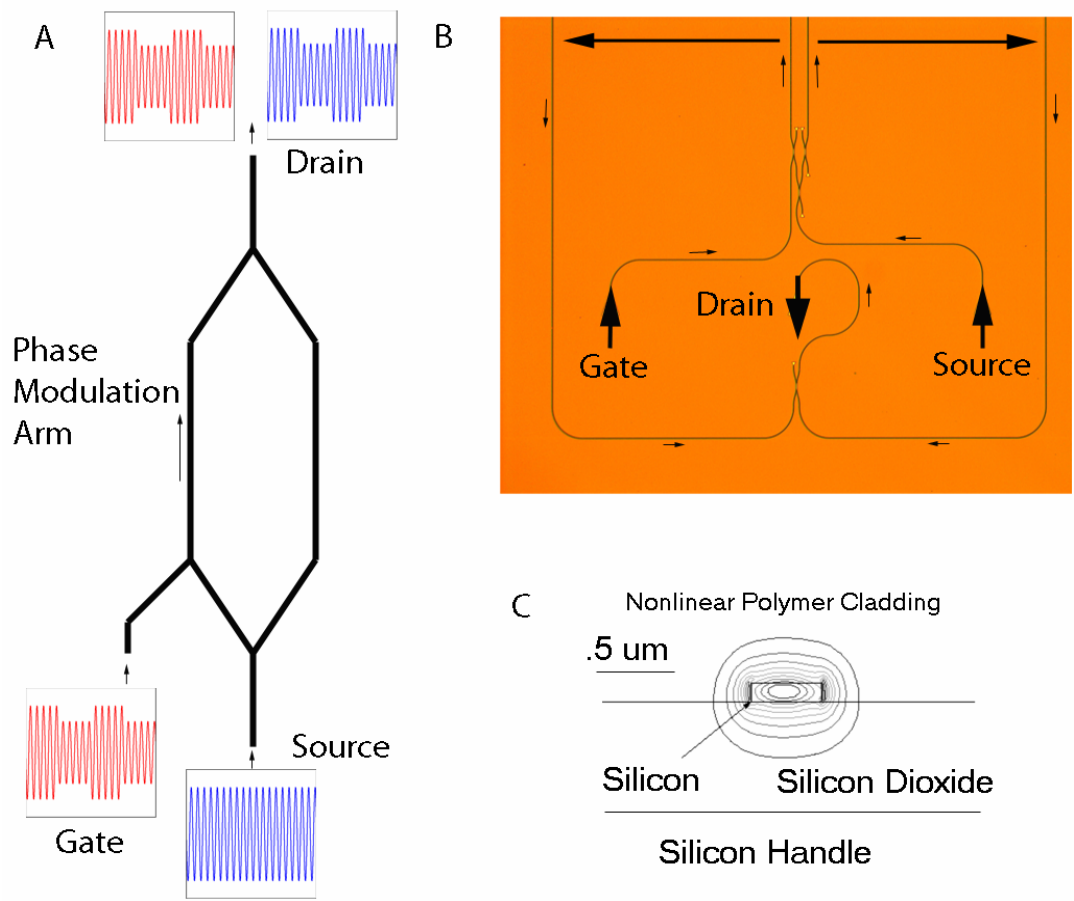

Figure 1: **Device Layout** Panel A shows a logical diagram of the operation of the high-speed all-optical modulator. The gate (red) signal has its intensity modulation transferred to the source (blue) signal via nonlinear phase modulation in one arm of the Mach–Zehnder interferometer. Panel B shows an optical-microscope image of part of the actual device. Panel C shows the mode pattern of the optical signal in the waveguide. Contours are drawn in 10-percent increments of power.

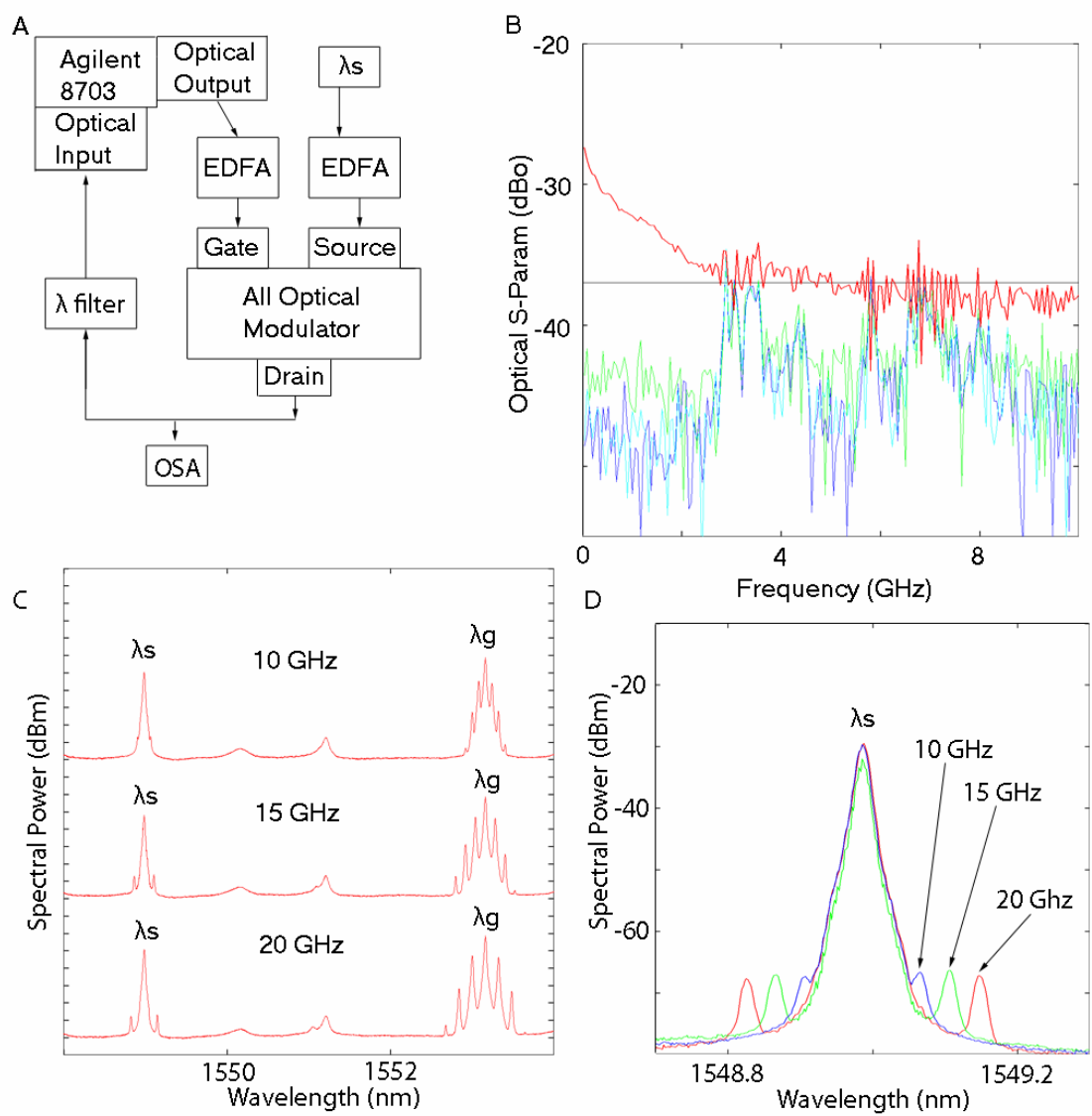

Figure 2: **Gigahertz Modulation Data** Panel A shows the logical diagram of the experiment to measure the optical S parameter. Panel B shows the measured S parameter for the device in various circumstances. The red curve is the measured value of the S parameter when both the gate and source lasers are on. The control measurements — taken when the signal laser is off, when the pump is off, and when all lasers are off — are shown by the green, blue, and teal curves, respectively. Based on the spectral four-wave –mixing data shown in Figure 3, we can calculate that the optical S parameter associated with the ultrafast optical measurementsis –37 dB. This level is shown in panel B as a black line. Clearly, the 10 GHz S-parameter and the calculated S-parameter from the terahertz data are in close agreement. Panels C and D show optical-spectrum traces taken for various sinusoidal radio-frequency–intensity modulations on the gate. The intensity modulation of the gate laser results in sidebands on the output, located near the source wavelength at the appropriate locations for each input-modulation frequency. Panel D shows a detail of the device output near the source wavelength for modulation at 10 GHz, 15 GHz, and 20 GHz.

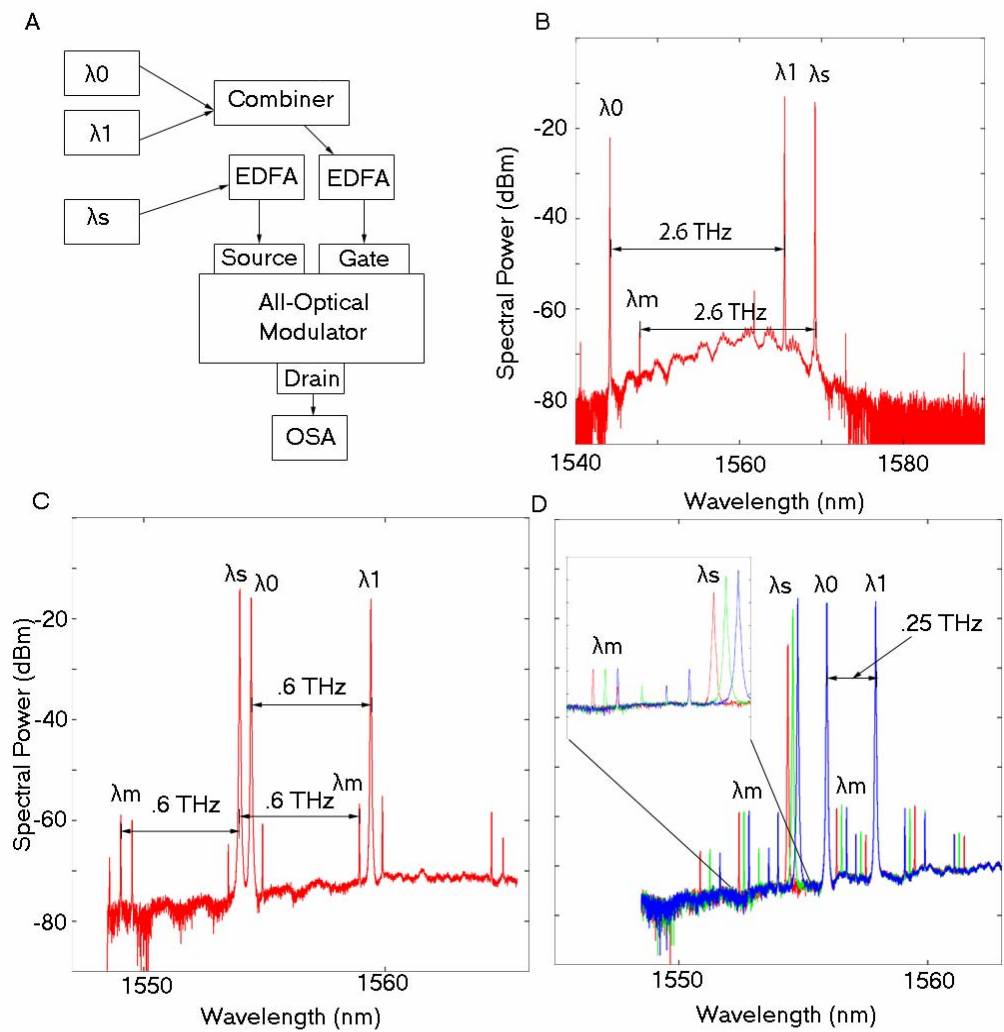

Figure 3: **Four-Wave–Mixing Experiments** Panel A shows the logical diagram of the experiment. Panels B, C, and D show the spectral output in the optical-spectrum analyzer for gate-laser spacings of 2.6 THz, 0.6 THz and 0.25 THz, respectively. The gate lasers are labelled λ0, λ1; the source signal is labelled λs. For the 2.6 THz plot, the relevant wavelength values are  λ0=1544.3 nm, λ1=1565.6 nm, and λs=1569.3 nm. The primary four-wave mixing sideband of the source signal is labelled λm.  Several other four-wave mixing peaks are visible. In addition, an inset in panel D shows the detail of the source and one set of sidebands as the former is tuned in increments of 0.2

nm. The change in the intensity of the central peak as the source wavelength is tuned, but not of the sidebands, is characteristic of the Mach–Zehnder interferometer's behavior.

**Methods**

This section describes the mathematical methods and derivations referenced in the body of this work.

**Nonlinear Phase-Modulation Mechanism**

To understand the limitations of the nonlinear phase-modulation mechanism, we must consider both the waveguide loss and the dispersion of our waveguides. Together, these two parameters establish a limitation on the amount of phase shift that can be obtained, or the rate of change of the phase shift that can occur. Our waveguide loss has been extensively calibrated through other test structures, and is known to be −7 dB/cm. The effective index of these waveguides near 1550 nm is well modelled by

$$n_{eff}(\lambda) = 1.93 - 1.21 * (\frac{\lambda(nm) - 1550}{1550}) + .934 * (\frac{\lambda(nm) - 1550}{1550})^2$$

As noted, this expression corresponds to a group velocity dispersion of about −2000 ps nm$^{-1}$km$^{-1}$, a value typically encountered for such waveguides. Note that, unlike a device dependant on four-wave–mixing conversion, in our device the momentum mismatch noted in Equation 2 does not affect the phase shift experienced by the source wavelength. Thus, the only limitation on the bandwidth of the device is the tendency of a pulse to scatter based on dispersion, and the tendency of pulses on different wavelengths to travel at different velocities. It is straightforward, using Fourier analysis, to analyze the time-domain behaviour of such pulses in the waveguide. The pulses used in the simulation shown in Figure 4 (supporting materials) were constrained to fit within 0.5 THz of the carrier signal, and the carrier signals were separated by 1.1 THz, ensuring that both pulses could propagate in the same waveguide and remain distinct.

**Derivation of Formula for Phase Shift**

This section provides a justification for the expression relating the fractional wavelength conversion to the phase shift.

We begin with the standard equation of nonlinear optics, modified slightly for the dispersive nature of the waveguide. Here, χ3 is in (m/V)$^2$.

$$\left(\frac{\partial^2}{\partial z^2} - \frac{1}{(c/neff)^2}\frac{\partial}{\partial t^2}\right)Ex(z,t) = \frac{1}{c^2}\frac{\partial^2}{\partial t^2}(\chi^3 Ex^3)$$

Here, Ex is the collection of all modes:

$$Ex = \sum a_i \exp(ik_i z - i\omega_i t) + a_i * \exp(-ik_i z + i\omega_i t)$$

Note that Ex is always real valued, as is required for use with the equations of nonlinear optics. Using the standard slowly varying amplitude approximation, all the linear terms are assumed to vanish for propagating modes, except for

$$\sum 2ik_i \frac{\partial a_i}{\partial z}\exp(ik_i z - i\omega_i t) + c.c. = \frac{\chi^3}{c^2}\frac{\partial^2}{\partial t^2} Ex^3$$

How we proceed depends on the particular nonlinear process under study. In the case of energy conversion by four-wave mixing, taking $\omega_2$ as the generated frequency, and $\omega_1$, $\omega_0$ as the pump beams, when $\omega_2=2\omega_1-\omega_0$, we have

$$2ik_2 \frac{\partial a_2}{\partial z} \exp(ik_2 z) = \frac{\chi^3}{c^2} 3(-i\omega_2)^2 a_1^2 a^*_0 \exp(2ik_1 z - ik_0 z)$$

$$a_2(L) = \frac{i3\omega_2^2 \chi^3}{2c^2 k_2} a_1^2 a^*_0 \int_0^L \exp(i(2k_1 - k_0 - k_2)z)dz$$

The final term, is simply a measure of the momentum mismatch among the three waves. With the dispersion roughly known for our waveguides from simulation and experiment, at scales around 1 cm length, and for frequencies less than 10 THz, this mismatched value is not significant, and the integral is nearly equal to L. This could also be deduced from the empirically observed fact that the four-wave mixing conversion efficiency did not vary substantially as a function of the magnitude of the difference frequency, at least for differences up to 1 THz.

Under this assumption ($|a_1|=|a_0|$), and under the condition that the two laser beams are nearly equal in power, the experimentally measured $f$ fraction is then

$$f = \frac{|a_2(L)|^2}{|a_1|^2} = \left(\frac{3\omega_2^2 \chi^3}{2c^2 k_2}\right)^2 L^2 |a_1|^4$$

In the case where one beam ($\omega_2$) is intensity modulated by another beam ($\omega_1$), the process can be written as

$$2ik_2 \frac{\partial a_2}{\partial z} \exp(ik_2 z) = \frac{\chi^3}{c^2} 6(-i\omega_2)^2 \, |a_1|^2 \, a_2 \exp(ik_2 z)$$

$$a_2(L) = a_2 \exp\left(\frac{i 6\omega_2^2 \chi^3}{2c^2 k_2} L |a_1|^2 \right)$$

Thus it turns out that, as seen by beam 2, propagation amounts to multiplication by a complex value with modulus 1. The absolute value of the argument of this value is readily seen to be twice f. That is,

$$\Delta\phi = |\arg(\exp\left(\frac{i 6\omega_2^2 \chi^3}{2c^2 k_2}\right) L |a_1|^2 ))| = 2\left(\frac{i 3\omega_2^2 \chi^3}{2c^2 k_2}\right) L |a_1|^2 = 2\sqrt{f}$$

Finally, note that the expression used in equation 4 was twice this amount: $4\sqrt{f}$. The value is doubled because the measured $f$ is artificially suppressed by a factor of ¼ due to the light passing through the final 3-dB coupler. The produced sideband is reduced by 3 dB, while the signal laser, because we biased the Mach–Zehnder modulator at max, is magnified by 3 dB.

**Modelling of Pulses in the Time Domain**

It is straightforward to analyze numerically the dispersion of a time-domain pulse in a dispersive waveguide; in Figure 4, we use the model for the refractive index of our waveguide to predict the time-domain behaviour of pulses at various speeds. The crucial question is whether the speed of a pulse at a gate wavelength will be sufficiently close to that of a pulse in the source wavelength to enable phase modulation by the nonlinear mechanism. In the simulations run, wavelengths at 1550 nm and 1541.2 nm were used. These wavelengths have a separation of 1.1 THz, ensuring that they can be separated spectrally without destruction of the signal being carried. For a device length of 1 cm, even for pulses of 1 ps in width — that is, terahertz pulses — the pulses are largely unseparated. Our calculations probably give a conservative estimate of the pulse spreading; the waveguide loss in the real devices ensures that the majority of the phase-shift transfer will occur in the first 0.5 cm of waveguide, whereas this simulation does not take that loss into account

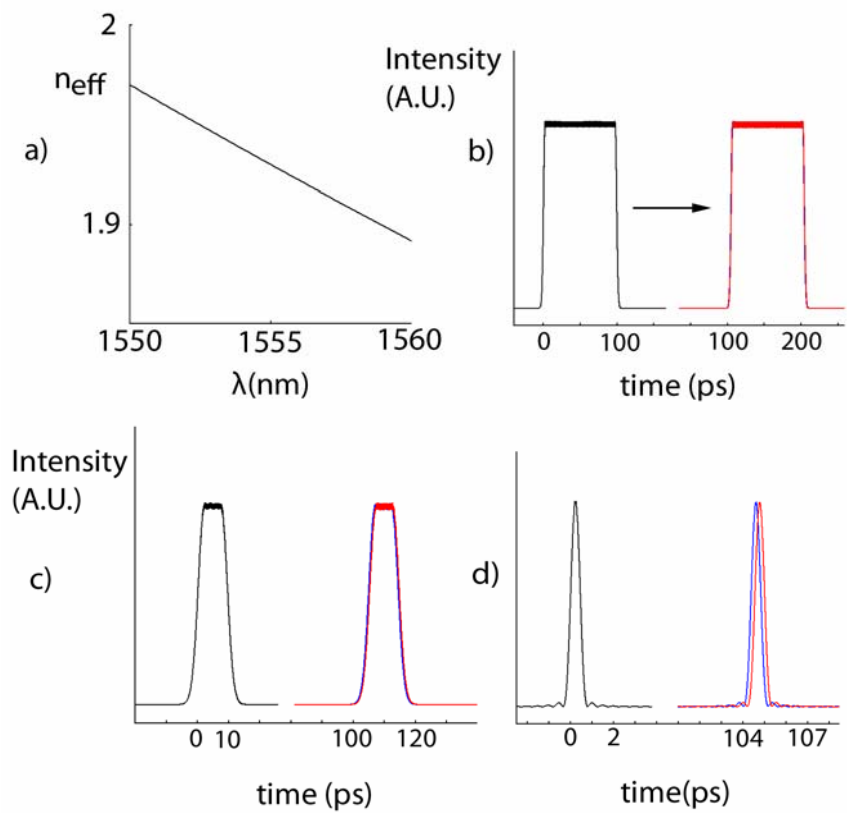

Figure 4: **Simulation of ultrafast time-domain behaviour** Panel A shows the effective index as a function of wavelength, a value that is slightly concave upward, leading to a negative group-velocity dispersion of about −2000 ps nm$^{-1}$ km$^{-1}$. Panels B, C, and D display what a time-domain–intensity detector would show at the start and end of the 1-cm modulation regions, for various input pulses. The signal in red is 1550 nm; that in blue is at 1541.2 nm. For clarity, attenuation due to waveguide loss has been normalized out of the second set of pulses. No blue line can be seen in panel B, because the pulses are nearly on top of each other.